\documentclass[epj]{svjour}
\usepackage{graphics}
\RequirePackage[colorlinks,citecolor=blue,urlcolor=blue,linkcolor=blue]{hyperref}
\begin{document}
\title{Impact of EMC effect on  D meson modification factor in equilibrating QGP}
\author{J.~Sheibani\inst{1} \and K.~Javidan \inst{1} \and A.~Mirjalili\inst{2}\thanks{\emph a.mirjalili@yazd.ac.ir (corresponding author)}  \and R.~Gharaei \inst{3} \and Shahin Atashbar Tehrani \inst{4}
%
}                     
%
%
\institute{  Physics Department, Ferdowsi University of Mashhad, Mashhad, Iran  \and Department of Physics, Yazd University, Yazd, Iran \and Physics Department, Hakim Sabzevari University, Sabzevar, Iran \and School of Particles and Accelerators, Institute for Research in Fundamental Sciences (IPM), P.O.Box 19395-5531, Tehran, Iran}
\date{\today}
%
\abstract{
In this article we employ the nuclear EMC effect to extract the parton distribution functions (PDFs) inside the Lead (Pb) and Gold (Au) nuclei. Extracted PDFs are utilized to obtain the transverse momentum dependent (TMD) ones, using the computing codes like   Pythia 8 or MCFM-10. Through this procedure TMDPDFs for charm and bottom quarks in Au at $\sqrt{s_{NN}}=200\;GeV$, Pb at $\sqrt{s_{NN}}=2.76\;TeV$ and $\sqrt{s_{NN}}=5.02\;TeV$ are calculated. To evaluate the validity of results and investigate the influence of nuclear EMC effect, the numerated TMDs are used as input to estimate heavy quark modification factor $R_{AA}$ at transverse plane $P_T$. This observable is calculated through numerical solution of the Fokker-Planck equation. For this purpose we need to   extract the drag and diffusion coefficients, using the hard thermal loop correction. It is done in the frame work of the relativistic hydrodynamics up to the third order approximation of gradient expansion. The results are compared with same solutions when the input PFDs are considered inside the unbounded protons where the nuclear effect is not included. {{}The comparison indicates a significant improvement of computed $R_{AA}$ with available experimental data when the EMC effect is considered.}
%
%
}
\maketitle
\section{Introduction}\label{int}
In standard quantum chromodynamics (QCD) quarks are the fermionic components of hadrons (mesons and baryons) and the gluons are considered as the bosonic components of such particles. The gluons, as bosons, are the force carriers of the QCD color force, while the quarks by themselves are their fermionic matter counterparts.{{}From the QCD predictions, a quark-gluon plasma (QGP) phase is arising out that is containing deconfined quarks and gluons,} but at an extremely high temperatures and/or densities \cite {int1,int2}. This state of matter is mainly created through heavy ion collisions \cite {int3,int4} or melting baryonic matter into each other due to an extreme pressure (like in the core of high-density astronomical objects \cite {int5} ) into a state at which hadrons are freed from their strong interaction. At this stage, hadrons look like as free quarks and gluons that compose initial baryonic matter where dissolve into the QGP medium. Light quarks and gluons construct the QGP bath, where heavy quarks (HQs) travelling there while interacting with QGP medium. HQs are produced at early stages of hard collisions and thus they are valuable probe to study the whole space-time history of the deconfined medium.\\

It is not surprising that, we expect to observe footprint of HQ distribution function at their origin i.e. the nuclei. Heavy quarks lose their energy through radiative and collisional interactions during their propagation in the QGP,  dominating at high and low transverse momentum $P_T$, respectively.\\

To finding a better perceptive from the properties of HQ re-scattering in the QGP, one can investigate the time evolution of HQ distribution function in transverse momentum $P_T$ plane using the Fokker-Planck (FP) equation in a thermally evolving medium \cite {int6,RAA2,ev1}. HQ energy loss can be characterized by the nuclear modification factor $R_{AA}$, which is proportional to the ratio of the particle yield in nucleus-nucleus (A-A) to that in proton-proton (p-p) collisions \cite{int8}.\\

In general, heavy quark initial distributions are generated by employing Monte Carlo simulations over proton-proton (p-p) collisions \cite{int8a,int8b,int8c}. It  can also be determined by fitting experimental data of $D$ and $B$  meson spectra over proton-nucleus collisions \cite{int8d}. The internal structure of collided particles is not taken into account, through these methods. Indeed, we have to extract initial conditions directly from hard scattering of colliding nucleons. On the other hand, it should be compared with initial conditions which have been taken from p-p collision at the same energy.\\

{{}Here, we take into account a nuclear effect, arising out from properties of collided particles, called EMC effect which was first observed  in 1983 by European Muon Collaboration at CERN \cite{1983}. In this regard one can say that parton distribution functions  of nuclei are different from
	PDFs of free nucleon, which clearly indicates that the quark
	or gluon freedom inside the bound nucleon is influenced by the nuclear
	medium environment. The nuclear medium effect on quark or gluon
	distribution attracts a lot of interests on both the experimental
	and theoretical sides, particularly since the discovery of
	the EMC effect in the valence-dominant region.}
Thus, through the EMC effect, nucleons in the nuclei matter, add their impacts into experimental observables.  Considering  the EMC effect directly enable us to investigate the influence of selected ion type in heavy ion collision, which according to our best of knowledge, has not been done before and is studied here for the first time.\\

It is expected that the QGP at relativistic heavy ion collision (RHIC) and  large hadron collision (LHC) is likely to be out of chemical equilibrium at the beginning of its formation. Therefore, it is necessary to consider this issue in our calculations. The energy loss of HQs at early stages of their propagation in the equilibrating QGP is very different from what happens in equilibrated medium \cite {int8e}.{{} Motivated by above explanations, we examine the influence of EMC effect into the $R_{AA}$ at early stage of equilibrating QGP}.\\

This paper is organized as it follows: In the next section, Sec.$\ref{ana}$, we discuss briefly  how to calculate  the PDFs and then TMDs for heavy quarks, taking into account EMC effect by running the APFEL legacy combination. Evolution of produced HQs from the initial function of partons  is discussed in
Sec.$\ref{hq}$. Computing the nuclear modification factor $R_{AA}$ by considering our derived TMDs, in two cases (with and without EMC effect) is done in the Sec.$\ref{QGP}$ and finally the conclusion is given in Sec.$\ref{Con}$.

\section{From PDFs to transverse distribution of HQs: employing EMC effect}\label{ana}
In order to investigate the thermodynamic properties of quark-gluon plasma and to  solve the Fokker-Planck hydrodynamic equation for this medium,  one needs to some inputs. One of these inputs is initial transverse momentum distribution of HQs in the nuclei-nuclei collisions at a desired and specified center of mass (CM) energy. Here the Pb-Pb and Au-Au collisions are illustrated.\\

Before to explain the method which leads us finally to HQs distribution function, let us have a brief review on EMC effect which provides  the PDFS while the nuclear effect is imposed on them.  This  effect is the surprising observation that the cross section for deep inelastic scattering from an atomic nucleus is different from that of the same number of free protons and neutrons (collectively referred to as nucleons). From this observation, it can be inferred that the quark momentum distributions in nucleons bound inside nuclei are different from those of free nucleons. As we said before this effect was first observed in 1983 at CERN by the European Muon Collaboration \cite{1983}, hence  named ``EMC effect". It was unexpected, since the average binding energy of protons and neutrons inside nuclei is insignificant when compared to the energy transferred in deep inelastic scattering reactions that probe quark distributions. Determining the origin of the EMC effect is one of the challenging subject in the field of nuclear physics. The primary theoretical interpretation of the EMC effect in the region x $>$ 0.3 was simple: quarks in nuclei move throughout a larger confinement volume and, as the uncertainty principle implies, they carry less momentum than quarks in free nucleons. The reduction of PDF ratio in nucleus with respect to unbounded nucleon  at lower x, named the shadowing region, was attributed either to the hadronic structure of the photon or, equivalently, to the overlap in the longitudinal direction of small-x partons from different nuclei. These notions gave rise to a host of models: bound nucleons are larger than free ones; quarks in nuclei move in quark bags with 6, 9 and even up to 3A quarks, where A is the total number of nucleons. More conventional explanations, such as the influence of nuclear binding, enhancement of pion-cloud effects and a nuclear pionic field, were successful in reproducing some of the nuclear deep-inelastic scattering data \cite{data1,data2}. More technical details about EMC effect can be found in \cite{EMC1,EMC2,EMC3,EMC4,EMC5,EMC6,Khanpour:2020zyu}.\\

\begin{figure*}[htp]
	\resizebox{0.9\textwidth}{!}{%
		\includegraphics{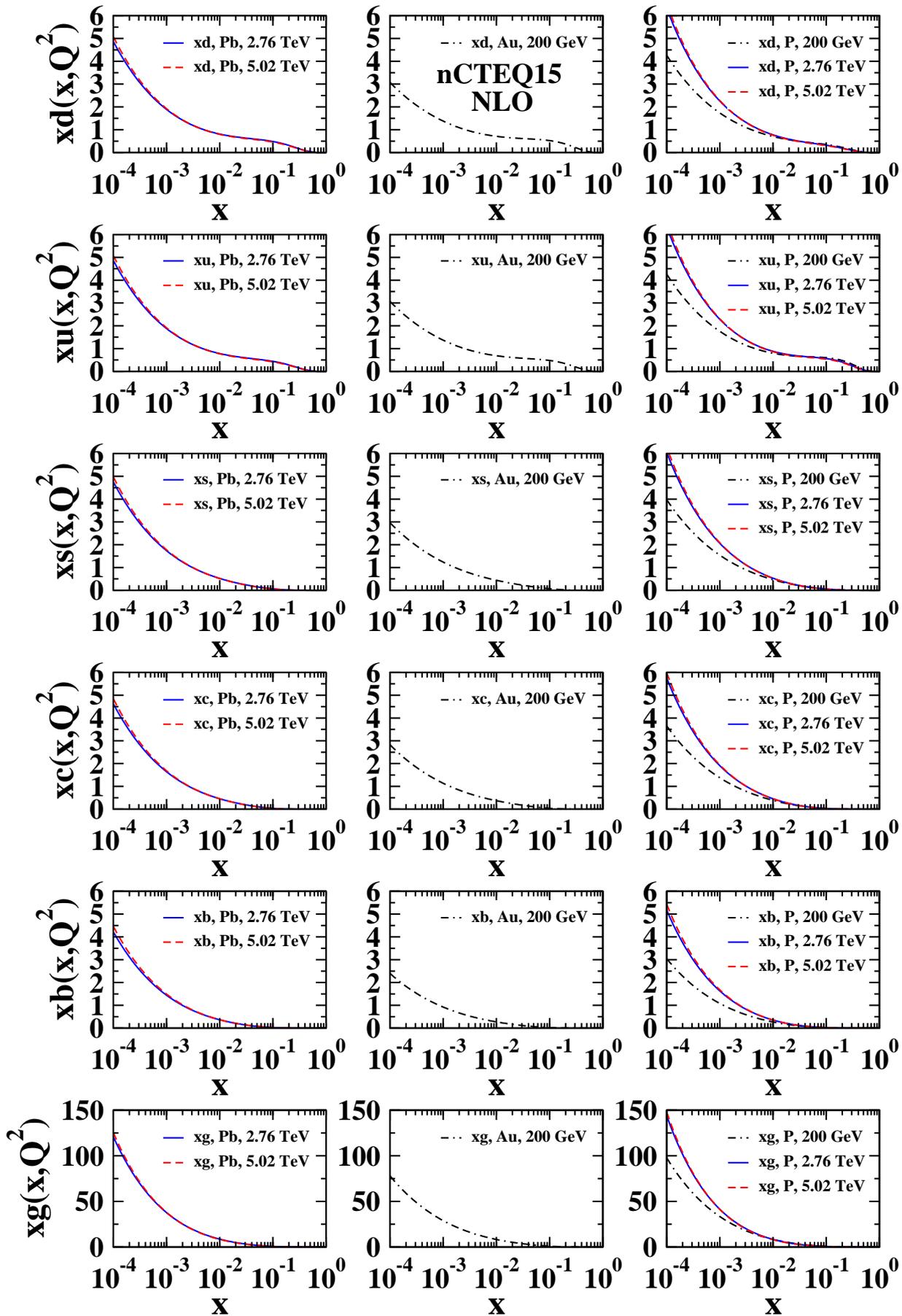}}
	\caption{\footnotesize (Color online) The $x$ dependent of bounded parton distribution for  Au  at center of mass energy  $\sqrt{s}=200$$\;$$GeV$ and for Pb at $\sqrt{s}=2.76,\; 5.02$$\;$$TeV$ energies and finally for the free parton distribution (right column) at $\sqrt{s}=200 \; GeV,\;$$2.76 \; TeV$,$\;$$5.02 \; TeV$ resulted from APFEL legacy  at the NLO approximation of  nCTEQ15 parametrization model~\cite{nCTEQ15}. \label{EMCPDF} }
	\label{xdependent}
\end{figure*}

{{} Now we back to our main aim.} To add the EMC effect into the HQ distribution functions, we  start from the early stage in obtaining the parton distribution functions. Thus, we are able to observe this effect more analytically. For this purpose the required distributions are achieved by running  the APFEL legacy \cite{APFEL}, using nCTEQ15 model while the EMC effect is accompanied there  to yield us the bounded quarks inside the nuclei as a function of $x$-Bjorken variable at specified center of mass (CM) energy. One can obtain from APFEL numerical outputs, via a fitting procedure, parameterized functions for the bounded PDFs which are dependent on $x$ variable.
The results for the bounded parton densities including charm and bottom quarks are depicted in Figure \ref{xdependent}.
\begin{figure*}[htp]
	\resizebox{0.75\textwidth}{!}{
		\includegraphics{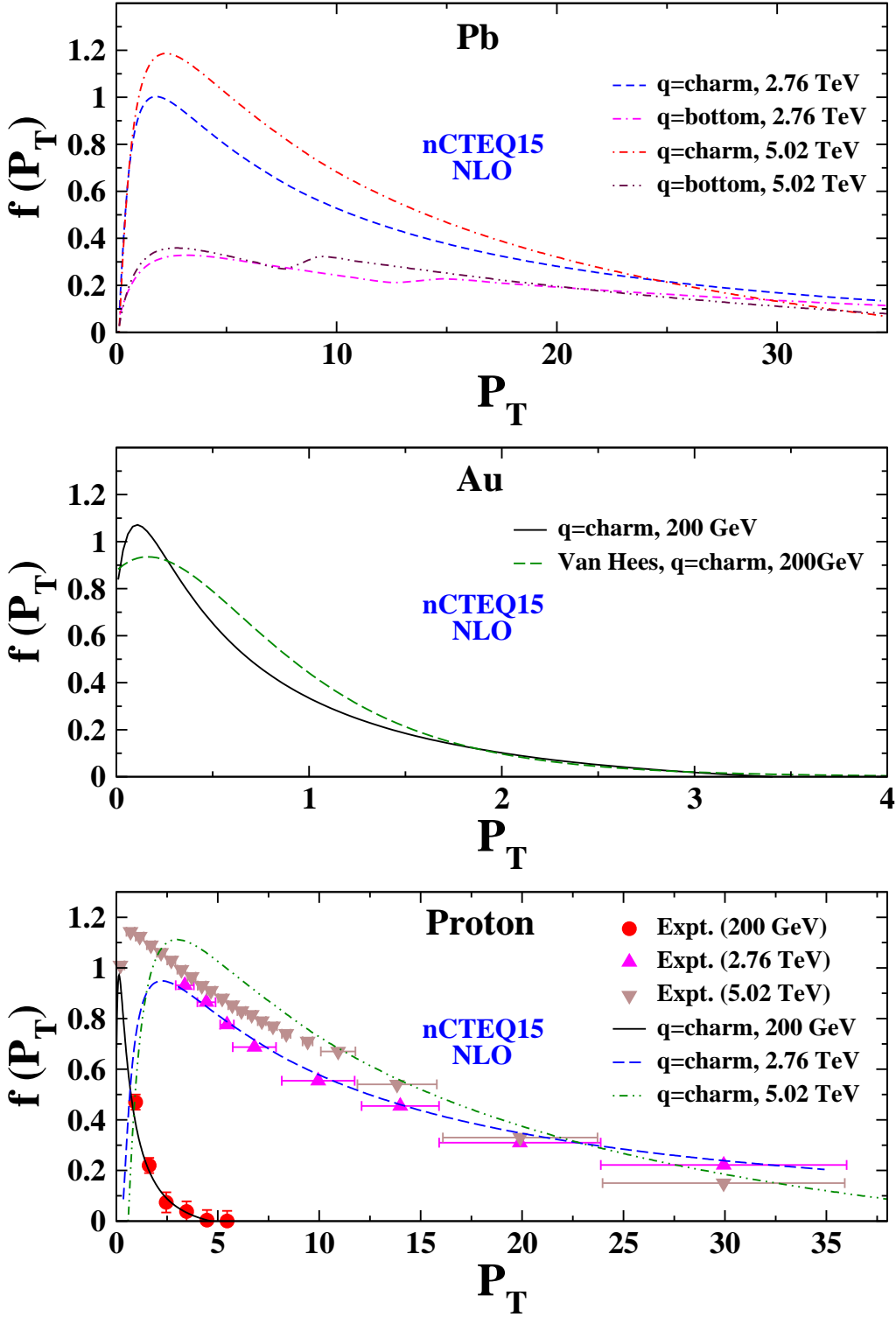}}
	\caption{\footnotesize (Color online) {{}The TMDPDFs for heavy $c$ and $b$ quark distributions} as a function of $P_T$ at early stages of  the GQP formation for Pb at center of mass energies $\sqrt{s}=2.76\;TeV$ and $\sqrt{s}=5.02$$\;$$TeV$ (upper panel), for Au at $\sqrt{s}=200$$\;$$GeV$ (central panel), and for proton at all center of mass energies (lower panel). TMDPDF distribution for Au at $\sqrt{s}=200$$\;$$GeV$ is also quoted from \cite{RAA2,ev1,int8a} and  included at middle panel. Experimental data for pp collision have been collected from \cite{p276} }
	\label{fig2}
\end{figure*}

Now, one can compute initial spectra of HQs in p-p, Au-Au and Pb-Pb collisions at desired center of mass energies.
There are several simulation codes with adjustable (or embedded) initial PDFs to extract spectra of HQs for above mentioned collisions. We employe the Pythia 8 \cite{pyt}  program to calculate transverse distribution of needed HQs. {{} This can  be done also by MCFM-10 program \cite{mcfm}.  The complete procedure which leads to transverse-momentum-dependent PDF contains three steps as they follow:\\
	
	(a) {{}At first we choose the grid data set of PDFs in the APFEL legacy program, arising out from many processes including  lepton-nucleon DIS, based on nCTEQ15 model.} By running it we achieve the outputs which are different nucleon PDFs at desired energy scales. These outputs are utilized as the input to do  simulation for proton-proton and nucleus-nucleus collision  that is describing at next step.\\
	
	(b) At this step we insert the extracted nucleon PDFs in preferably Pythia8 simulation code to find quark distributions after collision. The outputs of Pythia8 are not still transverse-momentum-dependent PDF but are parton distributions which, after collision, are produced inside proton or typical nucleus.\\
	
	(c)  Final step which lead us to the PDFs in the transverse plane is based on using the relation
	\begin{equation} f(x,P_T)=f(x)\frac{1}{(\pi <P_T^2>)} e^{\frac{-P_T^2}{<P_T^2>)}} \end{equation}
	which is a common procedure that can be seen in \cite{td2,td3}. The TMDs that only depend on $P_T$, is coming  by  integrating of $f(x,P_T )$ over $x$ variable.}\\

{{}Since we could compute the initial PDFs with and without EMC effect,} we are able to set these effects ``ON'' or ``OFF'' for all (or parts) of partons. {{} For this purpose the required procedure to yield us the PDFs that contain the EMC effect, are  including the following steps:\\
	
	I- We choose firstly data grid sets which involve PDFs inside specific nucleus like lead or gold at different energy scales.  These grids are providing by some models which in our case is again nCTEQ15 model. Using  these grid and running  the APFEL legacy program  the outputs would be different PDFs in the specific nucleus at desired energy scales which obviously are containing the  EMC effect as we plotted them in Fig.1 of our manuscript.\\
	
	II- To obtain the PDFs inside the free proton the above procedure is repeating, described in (a) step,  while we use  from the PDFs grid data inside the free proton, instead of nucleus,  as the  input for APFEL legacy. In this case the output would be the PDFs at desired energy scales which do not contain the EMC effect. If we inserted these PDFs as input into Pythia8 simulation code, the outputs would be PDFs inside specific nucleus after collision which obviously do not contain the EMC effect.\\
	
In summery from the running of  Pythia8 code, depending on the chosen grid data sets from nCTEQ15 model, we will get  two different sets of PDFs inside the specific nucleus after collision. One of them involves the EMC effect and the other one does not contain this effect. Following that the transverse-momentum-dependent PDF are achieved inside nucleus, as described in  (c), that depending on the used PDF types, containing or not containing the EMC effect respectively.}\\

{{} Fig.\ref{fig2} demonstrates the best results of TMDPDFs for  HQs at required center of mass energies. The calculated outcomes for proton-proton collision are compared with published experimental results at the same center of mass energies (in the lower panel of the Fig. 2) to control their validity. It may be noted that we have to normalize our results to reduce the numerical errors due to working with very large or very small numbers of data. Incompatibility in lower transverse momentum (in the form of a tiny shift) occurs due to the less accuracy in calculating the TMDs in lower $P_Ts$.}\\

Now we could manage several simulation setups with different initial PDFs. Once we set ``OFF'' the EMC effect for the charm quark {{} as} initial PDFs, while it was ``ON'' for other partons. Then, we compare outcomes with similar simulations but with including EMC effect for all partons. Results indicate considerable impact of initial HQPDFs on final HQ distributions due to hard collisions of collided particles. Finding reliable results need a plenty of simulations to reduce the uncertainty due to statistical errors which would be done in further works. Anyway, it might be an interesting study.\

\section{HQ Energy Loss in Equilibrating Quark-Gluon Plasma}\label{hq}

It is expected that the created QGP at hadron colliders is out of its chemical equilibrium. Therefore, it is necessary to consider this point in the evolution of produced HQs from the initial fusion of partons, in an equilibrating quark-gluon plasma \cite {int8e,Baier}. \
This issue directly affects on the collisional energy loss of HQs while propagating throughout a QGP
fireball. It is parameterized in terms of the distribution functions $\lambda_q n_F$ and $\lambda_g n_B$, where $n_F$ (Fermi-Dirac distribution) and $n_B$ (Bose-Einstein distribution) are quarks and gluon distribution functions respectively. Main parameters $\lambda_q$ and $\lambda_g$ are the fugacity factors which describing chemical non-equilibrium for quarks and gluons \cite{int8e}. In the early stages of plasma phase, energy loss of a HQ is dominated by gluons through $Qg \rightarrow Qg$ scattering ($\lambda_q=0$). The $Qq \rightarrow Qq$ interactions come into account as QGP goes toward its equilibrating state and thus $\lambda_q \rightarrow 1$. In order to find better approximation for HQ energy loss, we consider evolution of fugacity factors which are presented at the Ref. \cite{int8e}. Through this method we are able to calculate energy loss from early stages of creating the QGP fireball. \\

The most critical point for finding the time evolution of HQ distribution function during the interaction with QGP is correctly calculating the drag force, acted on the HQ in the QGP, and also calculating  diffusion of HQs in the medium. These quantities are related to the rate of collision and radiation energy loss per unit distance of HQ path, while traveling through the QGP. Realistic results for these values should be derived by considering the thermal effects using a gauge invariant field theory without infrared divergency. Here, we employ the hard-thermal loop (HTL) approach to calculate the energy loss for heavy quarks. In this method, diagrams of higher order in the loop expansion which contribute to the same order of the coupling constant are systematically resumed into effective propagators and vertices. The effective vertices are formed by adding the hard-thermal loop to the bare vertex \cite{HTL1,HTL2} (for more details please see also \cite {HTL4}).\\

We ignore collisions between heavy quarks which is by far an acceptable  approximation. In order to check the validity of our outcomes for drag and diffusion coefficients, we  compare our results with extracted values through different methods \cite{HTL4,HTL5,HTL6}. It may be noted that, the QGP fireball gets cooling while it expands over time. On the other hand, the temperature is a critical scale which controls the QCD coupling. Thus, we  use a temperature dependent function for the running coupling $\alpha_s(T)$ as \cite{HTL1}:
\begin {equation} \label{als}
\alpha_s(T)=\frac{6\pi}{\left(33-2N_f\right)\ln\left(\frac{19T}{\Lambda_{\overline{MS}}}  \right) }\;,
\end {equation}
where number of active flavors in the QGP has been taken as $N_f=3$  and{{} the QCD cut off parameter taken as $\Lambda_{\overline{MS}}=80\; MeV$.}\\

We need to evolve the HQ distribution function until the QGP temperature reaches the freeze-out temperature $ T_C$. \
We have studied almost all presented relations which describe collisional and radiation energy loss for HQs in the QGP. Following relations for collisional energy loss which are used in our simulations, we can write \cite {Baier}:

\begin{equation} \label {dedx1}
	-\frac{dE}{dx} = \alpha_s \tilde{m_g}^2 \ln \left[0.92 \frac{\sqrt{ET}}{\tilde{m_g}}e^{\lambda_q N_f/(12\lambda_g +2 \lambda_q N_f)}  \right] \;,
\end{equation}
where $\alpha_s$ is the strong coupling constant, $E$ is the HQ energy while $T$ is the QGP temperature. Also $\tilde{m_g}^2=\frac{4 \pi \alpha_sT}{3} \left( \lambda_g+\lambda_q \frac{Nf}{6} \right)$ when $N_f$ as before is the number of active flavors. We  examine other presented relations for collisional energy loss. By considering equilibration treatments, following expression can be applied \cite {de}:\\

\begin{eqnarray} \label {dedx2}
	\nonumber
	& -\frac{dE}{dx}& = \frac{4 \pi \alpha_s^2 T^2}{3}\left[ (\lambda_g+\lambda_q \frac{N_f}{6}) \ln \left( \frac{ET}{\tilde{m_g}^2} \right)+\right. \\
	&&\left.\frac{2}{9}\ln\left( \frac{ET}{M^2} \right) +c(N_f) \right]\;,
	\end {eqnarray}
	in which $c(N_f)=0.146N_f+0.05$. {{} This expression is mainly accurate for $E \gg M$ where $M$ is the HQ mass. It should be mentioned that, this  and other required estimations are extracted by some considerable approximations. Thus we have to apply these relations together with some conservation rules.} \\
	
	The density of quarks and gluons change in proper time, while the QGP evolves toward its equilibrated situation. Variation of QGP constituents changes the Radiation energy loss of HQs as well. We also consider this issue in our calculations through adding needed changes in raditive energy loss relation which has been taken from Ref. \cite {de}. Fig.3 demonstrates collisional and radiation energy loss for charm quark at the early stages of QGP formation ($\lambda_q=0$) and after equilibration ($\lambda_q=1$). This figure shows that collisional energy loss increases {{} by rasing $\lambda_q=0$ to $\lambda_q=1$}  but radiation energy loss decreases during QGP equilibration. Changes in energy loss due to plasma equilibrating are not significant for low-energy HQs. But for high-energy particles, such effects are considerable. Because of nonuniform impacts (with respect to particle momentum) of equilibrating QGP on HQ energy loss, this issue will produce indelible effects on the nuclear modification factor $R_{AA}$ which we discuss about it later on.\\
	\begin{figure}[htp]
		\centering
		\resizebox{0.5\textwidth}{!}{
			\includegraphics{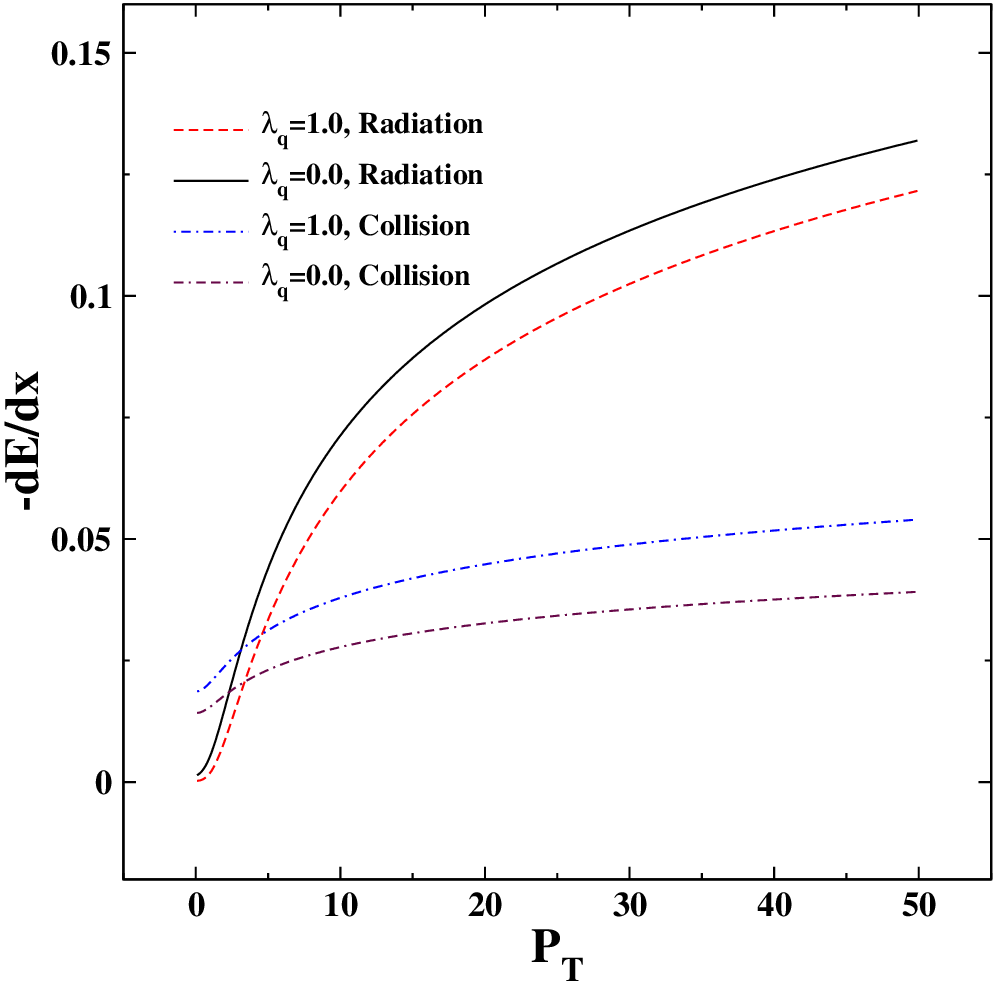}}
		\caption{\footnotesize (Color online) Collisional  and radiation energy loss of charm quark as a function of transervse momentum, $P_T, $at early stage of QGP formation ($\lambda_q=0$) and equilibrated QGP ($\lambda_q=1$) in a QGP with temperature T=0.4 GeV. }
		\label{energy loss}
	\end{figure}
Now we have introduced our key phenomena theoretically.  In the following, we have to examine impacts of these issues in time evolution of QGP observable. According to the Fig.\ref{fig2} and Eqs.(\ref{dedx1},\ref{dedx2}), we find that most of new effects can be studied by considering critical changes in quark distribution functions at transverse plane. This leads us to focus on calculating the nuclear modification factor $R_{AA}$ which is highly sensitive to HQ distribution functions as an initial condition. Therefore, we apply these treatments to calculate $R_{AA}$ for D mesons as a straightforward and clear process {{} which match properly} experimental data. Other observable like $\nu_2$ as an elliptic flow coefficient can also be inspected. Because of different nature of $R_{AA}$ and $\nu_2$, the nuclear modification factor is {{}an adequate  measuring tool} to quantify these effects.\\
	
We could now evolve the initial distribution of HQs from the early stages of QGP creation. It may be noted that, in previously presented works, evolution of HQ distributions have been calculated from initial proper time, mainly from $\tau_0=0.6 fm/c$ for RHIC and $\tau_0=0.3 fm/c$ at LHC, in Au-Au and also Pb-Pb collisions \cite{tc2} {{} which in our consideration, depending on the fitted results, would be changed.}

\section{Nuclear Modification Factor }\label{QGP}
	In this section we use our treated heavy quark distribution functions, extracted in previous section by considering the nuclear effects (at transverse plane) to evaluate a very famous observable in equilibrating quark-qloun plasmas. One of the most important phenomena related to the heavy ion collisions is the heavy quark energy loss while traveling through a quark-gluon plasma medium. This effect is quantified by the nuclear modification factor $R_{AA}$, which is defined as the ratio of the particle distribution (as a function of transverse momentum $P_T$) in nuclei (A-A) collisions $\frac{dN^{AA}}{dP_T}$, relative to the same spectrum in the proton-proton collisions, $\frac{dN^{PP}}{dP_T}$ as \cite{RAA1,RAA2}:
	\begin {equation} \label{RAA}
	R_{AA}(P_T)=\frac{1}{N_{coll}}\frac{\frac{dN^{AA}}{dP_T}}{\frac{dN^{PP}}{dP_T}}\;,
	\end {equation}
	where $N_{coll}$ is the average number of primordial binary nucleon-nucleon collisions for a specific centrality class. In the absence of any medium effects, $R_{AA}$ will be equal to one.\\
	
The nuclear modification (suppression) factor $R_{AA}(P_T)$  can also  be calculated by $\frac{f_f (P_T)}{f_i (P_T)}$, where $f_f (P_T)$ is the final transverse momentum distribution of heavy quark, after evolution due to traveling in the QGP medium \cite{ev1,ev2,ev3}. The $f_i (P_T)$ is related to initial spectrum before (or without) the evolution. It is clear that, the suppressed HQ distribution function for $A-A$ collision due to traveling in the QGP should be compared with HQ distribution function for the $p-p$ collision at the same energy.\\
	
{We know that if the  nuclear effect is not considered at all in our calculations then  the $R_{AA}$ ratio  in our  computations  would be equal to one. The effect of nuclear matter  can be considered in two ways. First by using the  Pythia 8 code  some features  of nuclear matter are taking into account and will  yield us  the HQ distribution  after A-A collision. This modification will be improved when we employ the extra EMC effect in our calculation.
		Fortunately, we could calculate the initial distribution with and without the nuclear EMC effect as we described it on page 5 in Sec.\ref{ana} . } Therefore, we are able to compare parton distribution function after collision with initial distribution function. It is interesting to note that, we can investigate impact of the nuclear effect for different types of partons separately.\\
	
{{}Getting the HQ distribution after collision,} to find the heavy quark (HQ) final state, $f_F (P_T)$, suppressed by the QGP, we need to calculate the time evolution of initial distribution of HQ ($f_{I} (P_T)$) in a thermal bath, using the Fokker-Planck (FP) equation \cite{FP1,FP2}:
	\begin {equation} \label{fp}
	\frac{\partial f(p,t)}{\partial t}=\frac{\partial}{\partial p_i}\left(pA_i(p) f(p,t)\right)+\frac{\partial^2}{\partial p_i \partial p_i}\left(D_{ij}(p) f(p,t)\right)\;,
	\end {equation}
	in which $f(p,t)$ is the HQ distribution function, $A_i(p)$ and $D_{ij}(p)$ are drag and diffusion coefficients in different spatial directions, {{} denoted by $i$ and $j$ indices}. As we  consider a spatially uniform QGP, the index $``i"$ can be eliminated. The QGP properties are encoded in drag and diffusion coefficients and indeed, in the time evolution of medium temperature. The drag force on the heavy quark can be calculated through $A=-\frac{dE}{dx}$ \cite{dd}. {{} One may calculate transport coefficients due to collisional and radiative energy loss separately and finally combine these quantities to find effective values for transport coefficients. In general, this method should be used with some cautions. Indeed, energy loss due to collisional process may provide considerable effects on the radiative process. Therefore, collisional and radiative energy loss should not be considered as independent processes \cite{de}.}  We have calculated the drag coefficient as $A(P_T)=A_{coll}(P_T)+ K A_{rad}(P_T)$, where $K$ is an unknown factor which should be fixed in an optimizing procedure. By considering the Einstein's relation {{} \cite{HTL4,FP2,dd}}, one can define the diffusion coefficient as $D=MTA$, where $E$ and $T$ are HQ mass and QGP temperature respectively \cite{dd,Das:2015ana,Akamatsu:2008ge}.\\
	
Eq.(\ref{fp}) can be solved, if we prepare three basic inputs: time evolution of the medium temperature through thermodynamical properties of the QGP, the drag and diffusion coefficients to fully describe the expanding QGP bath where HQs move therein and of course, initial transverse momentum distribution of considered HQ  \cite{FPS}. In the following we solve time evolution of HQ distribution functions by considering the EMC effect and without this effect. Thus we can compare the results to investigate the influence of EMC effect.\\
	
To describe the QGP as a fluid at higher energies, we need the relativistic version of hydrodynamics. The theory should be applicable for microscopic processes at higher momentum scales too. At high-energy fluid flows relevant for the LHC, it is expected that higher-order corrections of hydrodynamic approximation play an important role in the gradient expansion series. Thus, we  use the gradient expansion at third order in four dimensions to calculate time evolution of the QGP energy density in our work with respect to the proper time  \cite{3rd1}. Reminding that time evolution of the  {{} fluid  temperature} (extracted from the energy density) has been calculated at second order of approximation, and we are able to calculate the viscosity to entropy ratio $\frac{\eta}{s}$ as a function of proper time in a closed form ({{}see the Appendix)} \cite {3rd1,3rd2}.\\
	
As mentioned before, we  consider an equilibrating QGP. This means that we should calculate the HQ energy loss from early stage of the QGP formation, not after equilibrating QGP ($\tau_0$) as most of investigations have been done on this direction. For doing this, according to \cite{int8e} we  numerically solve time evolution equation for the fugacity factors $\lambda_q$ and $\lambda_g$ with $\lambda_q=0$ and unknown value of $\lambda_g$ as initial conditions. We  adjust the initial value for $\lambda_g$ by fitting the final result of $R_{AA}$ on experimental data.\\
	
{{} In continuation} calculated distribution function at temperature $T_C$ should be convoluted with the suitable fragmentation function to find the $P_T$ distribution function for $D$ meson. We employ the Peterson fragmentation function $D^D_c(z)=\frac{1}{z\left(z-\frac{1}{z}+\frac{\epsilon}{1-z} \right)^2}$ where $z =\frac{p_D}{p_c}$ is the momentum fraction of the D meson which is fragmented from the charm quark. We use $\epsilon=0.05$ in our numerical calculations \cite {fz1,fz2}. Such fragmentation function should be treated for using in the transverse plane \cite{td1}. The Gaussian model is widely used to include the transverse momentum dependence of fragmentation functions as:
	\begin {equation} \label {td1}
	D^D_c(z,P_T)=D^D_c(z)\frac{e^{-p^2_T/<P^2_T>}}{\pi <P^2_T>}
	\end {equation}
	where $<P^2_T>$ stands for the average transverse momentum of the produced D meson. We use $<p^2_T>= 0.20\;GeV^2$ for $p^2_T\le 1\;GeV^2$ and $<P^2_T>=0.25\; GeV^2$  for $P^2_T>1\;GeV^2$ \cite{td2,td3}.
	\newline
	\newline\\
	\begin{figure}[htp]
		\centering
		\resizebox{0.45\textwidth}{!}{
			\includegraphics{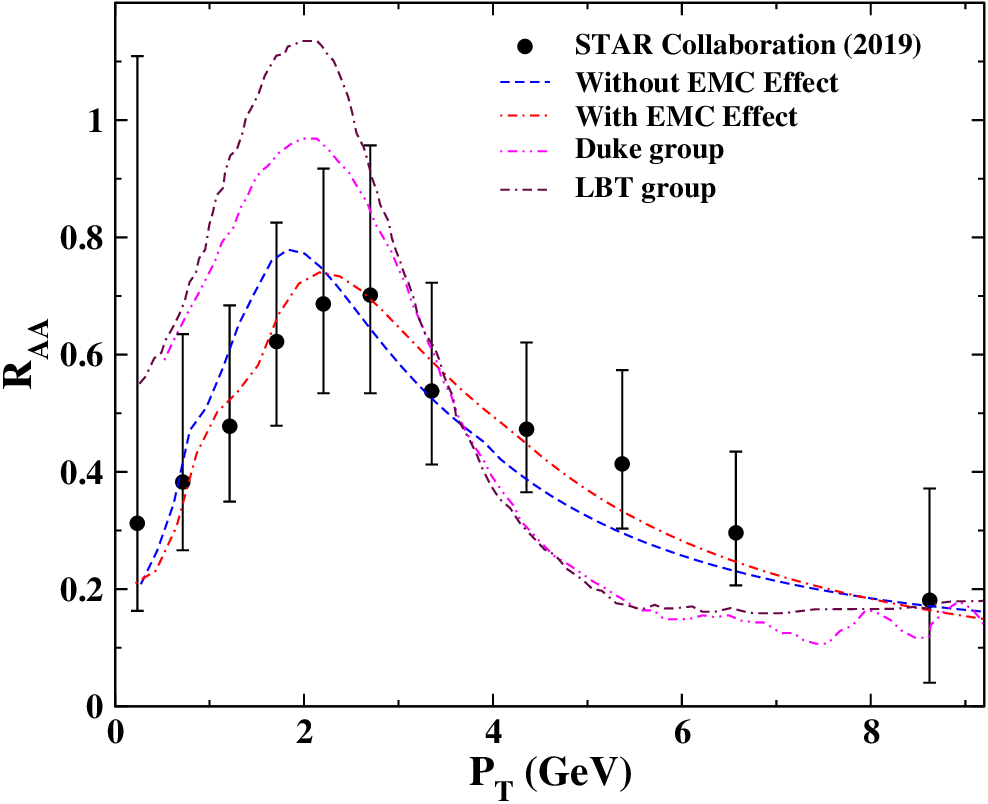}}
		\caption{\footnotesize
			(Color online) Nuclear modification factor of D mesons for central collisions at 200 $GeV$. The dashed line shows our result without the nuclear EMC effect. Dashed-dotted-dashed line shows our result by considering the EMC effect. The dotted-dashed-dashed-dotted  and dashed-dotted-dotted-dashed  lines represent the LBT and Duke  groups results~\cite{Duke1,LBT} respectively. Mid-rapidity experimental data are collected from STAR \cite{st200auau19}. }
		\label{raa200}
	\end{figure}\\

Now we are ready to examine our used method to calculate the HQ distribution function with respect to the transverse momentum $P_T$, containing the nuclear EMC effect. Procedure starts with initial distribution function of charm quark entering into the equilibrating (and expanding) QGP, with an unknown initial temperature $T_0$ at the center of fireball. As our distribution functions are calculated for $P_T>1\;GeV$, we  {{} extrapolate} results to find HQ distribution in $P_T \le 1\; GeV$. However, we  control our results in the region $P_T \le 1\;GeV$ by outcomes extracted using the presented method in \cite{td2,td3}, but our results are mainely valid for $P_T>1\;GeV$.\\
	
Evolution is simulated by getting numerical solution for the FP equation, at third order relativistic hydrodynamics until the fireball cools down into the freeze-out temperature. {We use thermal freeze-out to evaluate HQ distribution function in an expanding QGP. Thermal freeze-out is very dependent on the pressure gradients created in the fireball, from its central to the peripheral region. As the pressure gradient is a function of QGP energy density and dimension of fireball, the thermal freeze-out should be identified for different accelerators. We  use $T_c=155\;MeV$ for LHC \cite {tc1,tc2} and $T_c=170\;MeV$ for RHIC \cite {tc2,tc3}.}\\

{{} It may be noted that initial temperature should be calibrated against bulk matter observable. Indeed, this quantity is estimated by challenging measurement of direct photon, however model comparisons are needed because direct photons are also created at later stages of the heavy ion collisions. For solving the Fokker-Planck equation, we need initial temperature as an input with acceptable precision. Extracting this quantity through model dependent procedures only give us a range of values greater than the minimum required precision, needed for our purpose. Therefore, we  fix this critical input through our fitting procedure.}\\
	
In summary, we  optimize final result to fit on experimental data by adjusting initial parameters: $\lambda_g (\tau=0)$, proportion coefficient  for energy loss $``K"$ and also initial temperature of QGP ($T_0$), by minimizing the unweighted Chi-squared value:
	\begin {equation} \label {xi}
	\chi^2=\sum_i \frac{\left(R_{AA}^e(p_{T_i})-R_{AA}^c(p_{T_i})\right)^2}{\sigma^2_i}\;,
	\end {equation}
	in which $R_{AA}^e$ and $R_{AA}^c$ are experimental and calculated values of suppression factor and $\sigma$ is related experimental error \cite{tc3}. As we use unweighted initial distributions for HQs to calculate $\frac{f_F (P_T)}{f_I^{p-p} (P_T)}$, a scaled factor is also needed. In order to fully demonstrate impact of the EMC effect, we calculate the $R_{AA}(P_T)$ with and without considering EMS effects at initial distributions.\\
	\newline
	\begin{figure}[htp]
		\centering
		\resizebox{0.5\textwidth}{!}{
			\includegraphics{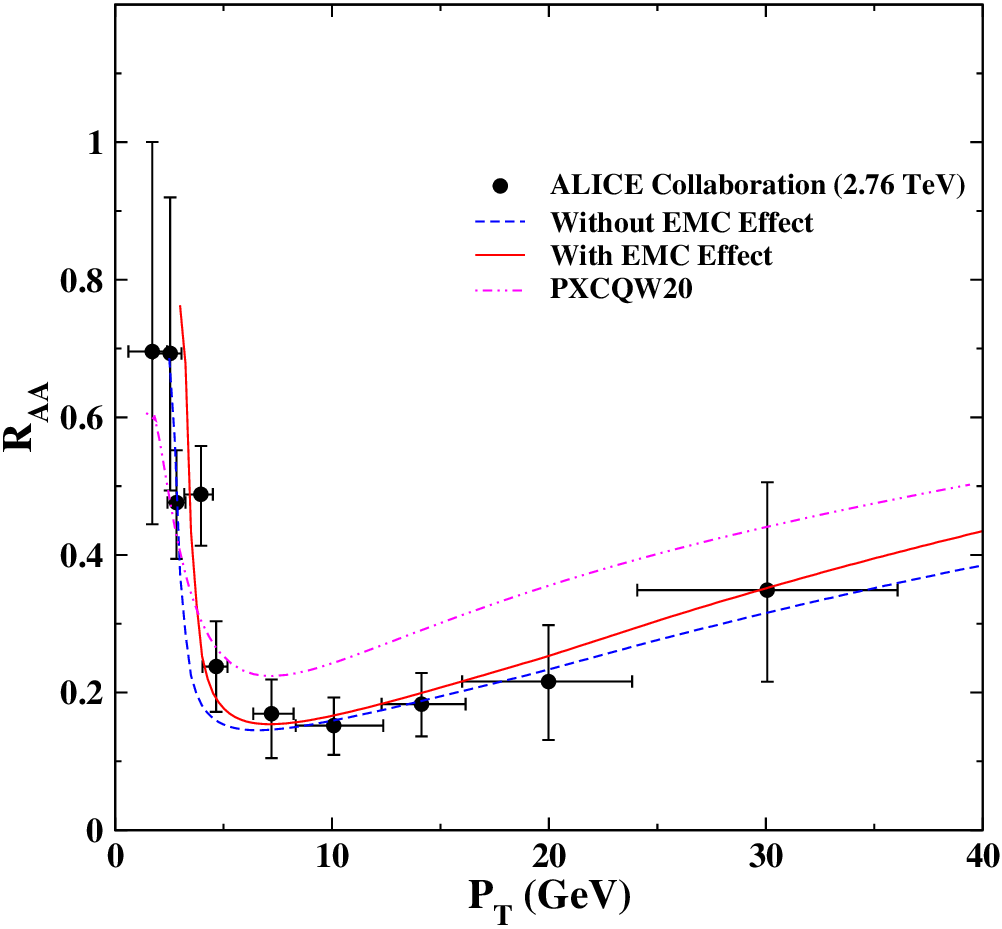}}
		\caption{\footnotesize
			(Color online) Nuclear modification factor of D meson for central collisions at 2.76 $TeV$. Solid line shows our result by considering the EMC effect, while dashed line represents our result without the EMC effects and dashed-dotted-dotted line presents the result of PXCQW20  group~\cite{Prado:2020imh}. Mid-rapidity experimental data are collected from ALICE collaboration \cite{2761,2762,2763}. }
		\label{raa276}
	\end{figure}
	
Fig.\ref{raa200} shows our results, arising our from calculating the $R_{AA} (P_T)$ for $D$ meson which involves charm quark. As mentioned before, the charm quark distribution is calculated for $P_T>1\;GeV$. Therefore, our results within $P_T < 1\;GeV$ region are not accurate. We  calculate the $\chi^2$ {value for data with $P_T > 1.5\;GeV$ to avoid errors due to unreliable region.} In the best fitting procedure, by using charm distribution function as initial condition with (without) considering the EMC effect, the minimum value of $\chi^2$ becomes 3.04 (10.57) by taking $\lambda_g=0.51, K=0.83, T_0=378\;MeV $ in both cases.\\

Fig.\ref{raa276} demonstrates our results for Pb-Pb at $\sqrt{s_{NN}}=2.76\;TeV$. Solid  line (dashed line) shows $R_{AA}(P_T)$ for D meson with (without) EMC effects. Best fitting results are extracted by considering $\lambda_g=0.41 (0.48), K=1.0, T_0=395\;MeV $ with (without) EMC effects.  Considering computed values, related to $P_T>3 GeV$ (far from unreliable region), the minimum values of $\chi^2$ in this procedure are 0.09 (3.25) for fitting with (without) EMC effect. Lower  $\chi^2$ values in this case (in comparison with results of $\sqrt{s_{NN}}=200\;GeV$) is because of larger experimental errors in experimental data at $\sqrt{s_{NN}}=2.76\;TeV$, so we can not compare values of $\chi^2$ for different experimental data sets.\\
	
\begin{figure}[htp]
		\centering
		\resizebox{0.5\textwidth}{!}{
			\includegraphics{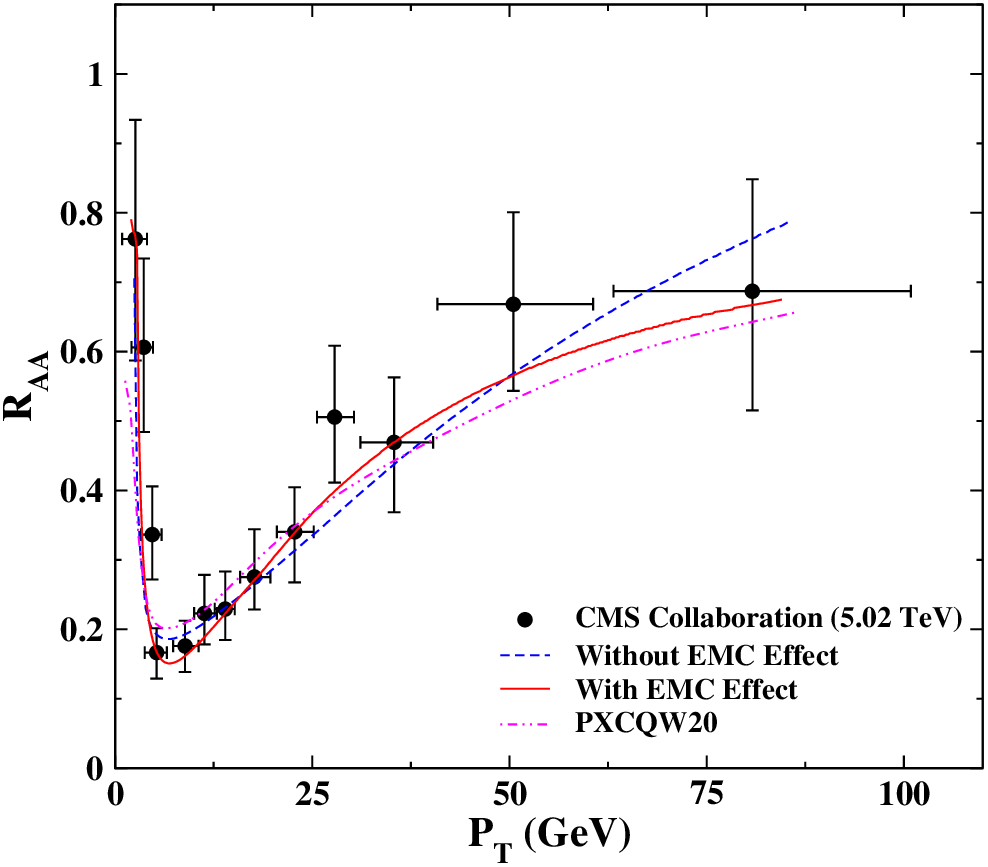}}
		\caption{\footnotesize
			(Color online) Nuclear modification factor of D meson for central collisions at 5.02 $TeV$. Solid line shows our result by considering the EMC effect, while dashed line represents our result without considering the EMC effect and dashed-dotted-dotted line presents the result of PXCQW20  group~\cite{Prado:2020imh}. Mid-rapidity experimental data are collected from CMS collaboration\cite{5021,5022,5023}. }
		\label{raa502}
	\end{figure}

Fig.\ref{raa502} shows results of calculations for $D$ meson modification factor $R_{AA}(P_T)$ in Pb-Pb collisions at $\sqrt{S_{NN}}=5.02\;TeV$.  The $\chi^2$ is minimized by taking $\lambda_g=0.38 (0.40) ,K=1.3$ and $T_0=403\;MeV$ with (without) considering the EMC nuclear effect. Its value becomes $\chi^2=0.02$ ($\chi^2=1.07$) for initial charm quark distribution with (without) EMC effect. This figure and the previous ones clearly indicate that EMC effect plays an important role in this procedure. But we should keep in mind that our method to calculate the HQ distribution as a function of $P_T$ is valid for $P_T>1 GeV$. Also, we do not consider some phenomena like coalescence in fragmentation process which is important in low energies \cite{coal}.\\
	
	\begin{table}
		\begin{center}
			\begin{tabular}{ c   c   c   c  c  c}
				\hline \hline
				$ \sqrt{S_{NN}} $ &  $\lambda_g$ & K & $T_0$ &  $\chi^2$ & EMC \\ \hline
				Au-Au 200 $GeV$ \hspace {1pt}\vline & 0.51 & 0.83 & 378 & 3.04 & Yes \\
				\hspace {67pt}       \vline	& 0.51 & 1.0 & 378 & 10.57 & No\\ \hline
				Pb-Pb 2.76 $TeV$  \vline &	0.41 & 1.0 & 395 & 0.09 & Yes \\
				\hspace {67pt}         \vline	& 0.48 & 1.0 & 395 & 3.25 & No\\ \hline
				Pb-Pb 5.02 $TeV$  \vline & 0.38 & 1.3 & 403 & 0.02 & Yes \\
				\hspace {67pt}         \vline	& 0.4 & 1.3 & 403 & 1.75 & No\\ \hline
			\end{tabular}
		\end{center}
		\caption{The best fitting results at different center of mass energies.  }
		\label{fit}
	\end{table}

Initial temperature and minimum $\chi^2$ for all fittings are presented in the Table.\ref{fit}. It may be noted that values of $\chi^2$ for a data set can not be compared with other ones. Since it is an unweighted value, the range of varying the $P_T$ and also experimental errors in each data set are very different. One can find from the Table.\ref{fit} that, considering the EMC effect does not change the best value of initial temperature.
	Our calculations are in good agreement with the experimental data. Most of the success in our numerical procedure is related to the calculation of the initial HQ distributions if compared with previously published results \cite {ev1,hf} (See Figure. 1 of \cite {hf1}). There are two noticeable points within results of  our calculations for the initial distribution at transverse plane. Considering computed values at lower transverse momenta lead to a smaller difference between $\chi^2$. This shows that the EMC effect creates its larger impact on higher $P_T$. It is an interesting result.\\
	
{{} As  can be seen from Fig.\ref {fig2}, using critical points of initial distribution functions, we can  find correctly the  minimum location of $R_{AA}(P_T)$ with respect to transverse momentum $P_T$ at high energy collisions.} We  also calculate the transverse momentum $P_T$ at which the suppression for bottom quark finds its maximum value. It is around $5 \le P_T \le 8.75\;GeV$ for $\sqrt{s_{NN}}=5.02\;TeV$  where uncertainties are due to lack of enough experimental data \cite {bquark}. Doing the required calculations, we predict that the minimum value of the bottom quark suppression for $\sqrt{s_{NN}}=2.76 TeV$ occurs at $4.75 \le P_T \le 7.25 GeV$.\\
	
{{} As Figs.\ref{raa200}, \ref{raa276} and \ref{raa502} clearly show, our procedure needs some considerations and improvements at lower $P_T$, however our calculated values for $R_{AA}$ at low $P_T$ are located in acceptable range of experimental data. As mentioned before, we  apply an {{} extrapolation} to find PDFs at low transverse momentums which provides some sorts of systematical errors at lower $P_T$ values. On the other hand, we do not include the D meson production from recombination.\\
		
It may be noted that we  use the D meson modification factor as a measure to demonstrate impacts of EMC effect in relativistic heavy ion collisions and our calculations in present situation clearly indicate that, nuclear effects create considerable impacts and should be taken into account at all center of mass energies.}\\
	
{{} To examine the validity of our results, we  compare outcomes with results reported in \cite{gr}. Our method to calculate the nuclear modification factor is somewhat different with numerical procedure in \cite{gr} but outcomes confirm each other. {{} Both results have some considerable remarks at low values of transverse momentum $P_T$ to estimate the $R_{AA}$.} According to findings of \cite{gr}: there are some problems in correctly calculating observable at low momentums, which might be raised from used {{} Landau-Pomeronchuk-Migdal (LPM)  model \cite{de}} and/or initial distribution functions. We  set  the LPM effects ``ON'' and ``OFF'' to compare the computed $R_{AA}$. Results show that differences due to ignoring the LPM is not so great and it cannot explain such differences at low $P_T$.\\

As our calculations clearly show, initial PDFs play a great role in final results of  observable system. One can observe such issue from Fig.\ref{xdependent} and Fig.\ref{fig2}. It seems that calculating nuclear impact through the EMC effect is a very good selection.}
	
\section{Conclusion}\label{Con}
	The nuclear suppression factor $R_{AA}$  for charm quarks in an equilibrated QGP as a function of transverse momentum $P_T$ has been calculated for Au-Au collision at $\sqrt{S_{N N}}=200 GeV$  energy as well as Pb-Pb collision at  $\sqrt{S_{N N}}=2.76 TeV$ and $\sqrt{S_{N N}}=5.02 TeV$ energies. To calculate this quantity in heavy ion collision one needs to transverse momentum dependence of heavy quarks. To increase the precision of calculations we imposed the nuclear effect on PDFs and attained the bounded PDFs. Influence of EMC effect on PDFs has been taken into account to calculate the transverse momentum dependence of heavy flavours created in the QGP after heavy ion collision. Time evolution of quarks and gluon densities in equilibrating QGP  have also been considered in calculating the heavy quarks modification (suppression) factor. Validation of calculated distribution functions and impacts of considering the EMC effect have been evaluated by applying results as input to calculate the $R_{AA}(P_T)$. Numerical calculations have shown that, {{} employing the EMC effect for calculating the initial conditions of the HQ-QGP interaction will improve the results, independent of used method to solve the  evolution equations of system.} The general form of extracted distribution functions at different center of mass energies are brightly in agreement with critical identifications of experimental outcomes.\\
	
We also  calculated the location of maximum suppression for the b quark at $\sqrt{s_{NN}}=5.02 GeV$  which is in agreement with available data. Following that we predicted that the maximum suppression for b quark has been occurred at $4.75 \sim 7.25 GeV$ for $\sqrt{s_{NN}}=2.76 GeV$.\\
	
Our study can be extended to include better models for the fragmentation functions at transverse plane. One further step is to  consider the effect of strong magnetic field generated in the QGP fire ball. In the following, collective effect due to the generating shock waves is also a challengeable research activity. We hope to report later on these issues  as our further research tasks.

\section*{Acknowledgments}
	J. S  and A.M acknowledge  Yazd university. First author, J. S, also thanks Ferdowsi university of Mashhad.  K.J. is thankful  Ferdowsi university of Mashhad.   R.G  is grateful Hakim Sabzevari university and  S.A.T  is appreciated school of particles and accelerators, institute for research in fundamental sciences (IPM). All authors  gratitude  their home universities  for the warm hospitality while this project has been performed. We finally appreciate R.Taghavi to help us to compute the primary TMDPDFs.

\section*{Data  availability statement}
The datasets generated during and/or analysed during the current study are available from the corresponding author on reasonable request.
	
\section*{Appendix}\label{App}
	We have used a fully theoretical procedure, avoiding pure Monte Carlo simulations to find physical sense about {{} the concerned  parameters in our analysis.} It is the reason to follow the presented methods in references \cite{de} and \cite{dd} to put effects of all phenomena on the table. Through these methods we can prepare a very controllable comparison between different results.
	
	To calculate the heavy quark distribution as a function of transverse momentum after traveling through the QGP, we have solved the Fokker-Planck equation, Eq.(\ref{fp}).
Initial conditions for numerical solution of this equation are: Heavy quark initial distribution function $f(P_T, \tau=\tau_0)$ as well as drag ($A(P_T, \tau)$) and diffusion coefficients ($D(P_T, \tau)$). Initial distribution function of heavy quark at transverse plane with and without including the EMC effect have been calculated numerically as explained in Sec.\ref{ana} {{} where the TMDPDFs,  plotted in Fig.\ref{fig2}, are including the EMC effect.} To calculate the drag coefficient by $A(P, \tau)=-\frac{dE}{dx}$ we need to energy loss of heavy quark while traveling into the QGP bath. Collisional energy loss for heavy quark in equilibrating QGP are given by Eqs.(\ref{dedx1},\ref{dedx2}). We have examined several proposed relations to calculate the radiative energy loss. According to our numerical simulations, the best applicable relation for radiative energy loss is as following \cite{rad1,rad2}:
	\begin {eqnarray} \label {rad}
	\frac{dE_{rad}}{dx}&=&24 \alpha_s^3(T) \rho_{QGP} \frac{1}{\mu_g} \left( 1-\beta_1 \right)\nonumber\\
	&\times&
	\left( \sqrt{\frac{1}{1-\beta_1}\ln \left( \frac{1}{\beta_1}\right)}-1  \right) F(\delta)\;, \nonumber \\
\end{eqnarray}
where
\begin {eqnarray}
F(\delta)&=&2\delta-\frac{1}{2}\ln \left( \frac{1+\frac{M^2}{s}e^{2\delta}}{1+\frac{M^2}{s}e^{-2\delta}}  \right)\nonumber\\
&-&
\left( \frac{\frac{M^2}{s}\sinh (2 \delta)}{1+2\frac{M^2}{s}\cosh (2\delta)+\frac{M^4}{s^2}} \right)\;, \nonumber \\
\delta &=& \frac{1}{2}\ln \left[ \frac{1}{1-\beta_1} \left( 1+\sqrt{1-\frac{1-\beta_1}{\ln \frac{1}{\beta_1}}} \right)^2  \right]\;, \nonumber \\
s&=&2E^2+2E\sqrt{E^2-M^2}-M^2,   \beta_1=\mu_g^2/(C E T)\;, \nonumber \\
C&=&\frac{3}{2}-\frac{M^2}{4ET}+\frac{M^4}{48 E^2 T^2 \beta_0}\nonumber\\
&\times&\ln \left[ \frac {M^2+6ET(1+\beta_0)}{M^2+6ET(1-\beta_0} \right]\;, \nonumber \\
\beta_0 &=&\sqrt{1-\frac{M^2}{E^2}}, \rho_{QGP}=\lambda_q \rho_q+\lambda_g \frac{9}{4}\rho_g\;, \nonumber \\
\rho_q &=& 16T^3\frac{1.202}{\pi^2},  \rho_g=9N_fT^3\frac{1.202}{\pi^2}\;,\nonumber \\
\mu_g &=& \sqrt{4 \pi \alpha_s  \left( \lambda_q+\lambda_g N_f/6 \right)}T\;. \label{many}
\end {eqnarray}
Here $M$ is the HQ mass, $T$ is the QGP temperature. {{} In above relations, the Landau-Pomeronchuk-Migdal (LPM) effects have been taken into account trough proper definition of the Debye screening for gluon. The LPM effects cause a reduction in scattering process between HQs and the QGP by adding a limit on the formation-time on the phase space of the emitted gluon in a way that the formation time, should be smaller than the interaction time \cite{de}.}

{{} Another important phenomenon is related to the namely dead-cone effect. The rate of radiative energy loss is proportional to the mean energy of the emitted gluons \cite{de}. The dead-cone effect includes  the radiative energy loss by adding a relation between mean energy of emitted gluon and gluon emission angle (for more details please see Sec.4 in \cite{dd}).}

{{} The effective drag coefficient should be obtained by adjusting the contribution rate of the collisional and the radiative energy loss, by fitting the theoretical calculations on the experimental data. The collisional energy loss due to interaction of HQs with light quarks is starting from the onset of fire-ball formation \cite{int8e,de}. For evaluating the contribution of collisional energy loss due to HQ-gluon interaction we have taken the $\lambda_g$ as fitting parameter. Thus, we need another fitting parameter to adjust the contribution of radiative energy loss.}{{} In this regard} effective drag coefficient is taken as $A(P_T, T)=\frac{dE_{coll}}{dx}+K\frac{dE_{rad}}{dx}$ where $K$ is free parameter that should be fixed in numerical calculations.\\

{{} We can consider the QGP as an ideal noninteracting fluid, that HQs are moving while interacting with plasma constituents. On the other hand, we assume that HQs do not interact with each other too. According to the Einstein relation, the diffusion coefficient is defined as               $D \propto \bar{\Delta}^2$, where $\bar{\Delta}^2$ is the mean square of the deviation between HQ and plasma particles velocities, in a given direction (at interaction interval $\tau$). In this situation, particles may move with a large velocity, but mean square deviation is considerably small and we can employ the nonrelativistic version of the Einstein relation \cite{D1,D2}.} Thus, we have used the relation $D(P_T,T)=MTA(P_T,T)$ to calculate the diffusion coefficient \cite{FP2}.
The relations in Eq.(\ref{many}) show that drag and diffusion coefficients are function of QGP temperature $T$, while {{} they are in fact function} of local time $\tau$ i.e. $T(\tau)$. This means that we need to some  proper equations to calculate time evolution of QGP temperature.\\

Time evolution of QGP parameters is derived from the conservation of the stress-energy tensor which is written in the gradient expansion of the velocity and temperature of fields. Time evolution of the QGP temperature at the zero-order approximation of gradient expansion is given by \cite{3rd,f1}:
\begin{eqnarray} \label{0}
	T(\tau)=T_0 \left( \frac{\tau_0}{\tau}\right)^{\frac{1}{3}}\;, \nonumber
\end{eqnarray}
in which $T_0$ and $\tau_0$ are initial temperature and initial local time respectively. We have taken reported value in published papers for initial time $\tau_0$ {{}\cite{tc2}}, while initial temperature has been taken as a free parameter {{} which is found by doing a  fit} over the experimental data.\\

Based on the consideration in \cite{3rd,f1} the first order relation for time evolution is given given by:
\begin{eqnarray}\label{1}
	T(\tau)= T_0 \left( \frac{\tau_0}{\tau}\right)^{\frac{1}{3}}\left[ 1+\frac{2}{3 T_0 \tau_0}\frac{\eta}{s}\left( 1-\left(\frac{\tau_0}{\tau}\right)^{\frac{2}{3}}\right)\right]\;, \nonumber
\end{eqnarray}
where $\frac{\eta}{s}$ is the viscosity to entropy ratio. In most of numerical simulations, this quantity has been taken as a free parameter which will be fixed through fitting the model on the experimental data. Implementing the viscous hydrodynamics for describing the heavy-ion collisions,
leads to a surprisingly small value of the shear viscosity to entropy density ratio, $\frac{\eta}{s}$. Proposed expressions for the time evolution of $\eta/s$ through the hydrodynamic modelling {{}which is containing high-energy heavy-ion collisions,} is very different and  fraught with many uncertainties.
We have examined several presented expressions for QGP $\eta/s$ in different papers \cite{ets1,ets2}. Below equation, provides acceptable values for this quantity:
\begin{eqnarray} \label {etas}
	\frac{\eta}{s}=\frac{3T^4}{4\epsilon}\frac{27.126}{g(T)^4 \ln \left(2.765g(T)^{-1}\right)}\;. \nonumber
\end{eqnarray}
Here $g(T)^2=4 \pi \alpha_s(T)$ and $\alpha_s$ is the strong coupling, given by Eq.(\ref{als}).\\

We have also used the presented expansion $\frac{\eta}{s}= 0.097+ 0.4955 (t-1)^2- 0.1781 (t-1)^3+ 0.01738(t-1)^4$, with $t=T/T_C$ \cite{ets3}. This relation is valid up to {{}$t=4.5$ Sec} , with quality of approximation $\chi ^2/dof.=0.0228$, which is acceptable for our calculations.\\

We have setup our calculations in radial coordinates in the transverse plane $r^2=x^2+y^2$ and $\tan \phi=y/x$, with proper time $\tau=\sqrt{t^2-z^2}$ and space-time rapidity $\Lambda=arctanh\frac{z}{t}$. By considering some conditions on the local equilibrium distribution, expanding the stress-energy tensor up to the third order term of expansion and using the fluid equation of state, the temperature $T$ relates to the fluid energy density $\epsilon$ and pressure $p$, as $\epsilon=3p=3T^4/\pi^2$ where {{} $\pi$, that is different from conventional mathematical symbol,}  can be found through fluid equations of motion \cite {3rd}. It may be noted that our main aim is to demonstrate the importance of adding the EMC effect  in which the initial condition is affected by it.{{} Therefore it is not harm for our results to use the ideal fluid equation of state.}

Following the set of equations  which  describes time evolution of the QGP variables up to the third order of expansions, one can write \cite{3rd,Amaresh1,Amaresh2}:

\begin{eqnarray} \label {em}
	\frac{d\epsilon}{d \tau}&=&-\frac{1}{\tau}\left( \frac{4}{3}\epsilon-\pi\right)\;, \nonumber \\
	\frac{d \pi}{d\tau}&=&-\frac{\pi}{\tau_ {\pi}}+\frac{1}{\tau}\left( \frac{4}{3}\beta_{\pi}-\lambda \pi-\kappa \frac{\pi^2}{\beta_{\pi}}\right)\;, \nonumber
\end{eqnarray}
where $\beta_{\pi}=\frac{4\epsilon}{15}$. The terms proportional to $\lambda=\frac{38}{21}$ and $\kappa=\frac{72}{245}$ are second- and third-order parts of approximation in the gradient expansion of the stress-energy tensor, respectively. The shear relaxation time $\tau_\pi$ is equal to the Boltzmann relaxation time $\tau_r$ \cite{3rd} so that: $\tau_{\pi}=\tau_{r}=5\frac{\eta}{s}\frac{1}{T}=\frac{\eta}{\beta_\pi}$, while $\eta = 187.129\frac{T^3}{g^4\ln g^{-1}}$ . Another relation to calculate the shear relaxation time is $\tau_{\pi}=\frac{2-\ln{2}}{2\pi}$ \cite{f3}.  As an acceptable approximation, we calculated the fireball temperature from energy density (calculated up to third order of approximation) using second order hydrodynamics relation \cite{f1}. Both results are very near to each other.\\

Now we have all needed equations to calculate time evolution of the {{} fluid temperature} at zero, to third order  approximation of dissipative hydrodynamics. The bulk viscous pressure vanishes by applying some conditions due to symmetries, and the shear stress tensor is fully specified by the difference between the longitudinal and transverse pressures. We have numerically solved equations using the $4^{th}$ order Runge-Kutta method. Time spacing grids should be chosen very carefully. At earlier time of evolution (which starts from initial time ($\tau_0$)), gradient terms find their maximum values. This means that, initial values of variables will reveal stronger deviations from ideal fluid dynamics. Therefore, we have to choose smaller time steps to solve the dissipative fluid dynamics equations at the beginning of numerical solution procedure. We have used the adaptive step-size method to control the numerical errors. Our calculations clearly show that, without such treatment, numerical errors become so large and results are meaningful.

\end{document}